\documentclass[a4paper,12pt]{article}
\usepackage[english]{babel}
\usepackage{amsmath,amsfonts,amssymb}
\usepackage{slashed,braket}
\usepackage{hyphenat}
\usepackage[numbers,comma,square,compress]{natbib}
\usepackage{feynmp-auto}
\usepackage{hyperref}
\usepackage[margin=28mm]{geometry}
\usepackage{enumerate}

\begin{document}
\begin{center}
\LARGE{Reciprocal of the CPT theorem}
\end{center}
\vspace{.2em}
\begin{center}
\textbf{Luis \'Alvarez-Gaum\'e\,$^{a,b}$, Moshe M. Chaichian\,$^{c,d}$,
Markku A. Oksanen\,$^{c,d}$,\\\vspace{.3em} and Anca Tureanu\,$^{c,d}$}
\end{center}
\begin{center}
$^{a}$\;{\it Simons Center for Geometry and Physics,\\
State University of New York Stony Brook,\\
NY–11794–3636, USA}\\
\vspace{.6em}
$^{b}$\;{\it Theory Department CERN,\\
CH-1211 Geneva 23, Switzerland}\\
\vspace{.6em}
$^{c}$\;{\it Department of Physics, University of Helsinki,\\
P.O.Box 64, FI-00014 University of Helsinki, Finland}\\
\vspace{.6em}
$^{d}$\;{\it Helsinki Institute of Physics,\\
P.O.Box 64, FI-00014 University of Helsinki, Finland}
\end{center}

\vspace{.2em}

\begin{abstract}
The CPT theorem originally proven by L\"uders and Pauli ensures the
equality of masses, lifetimes, magnetic moments and cross sections of
any particle and its antiparticle. We show that in a Lorentz invariant
quantum field theory described by its Lagrangian, CPT-violating
interaction alone does not split the masses of an elementary particle
and its antiparticle but breaks only the equality of lifetimes, magnetic
moments and cross sections. However, CPT violation in the mass term of a
field in the Lagrangian, which can be attributed to be due to the size
of the particle described by a form factor, breaks only the equality of
masses. Also it is shown that the two separate effects of CPT violation
in the interaction terms or in the mass term do not mix due to higher
quantum corrections and remain distinguishable. Thus, we urge the
experimentalists to search for such observable effects concerning
differences in the masses, magnetic moments, lifetimes and cross
sections between the elementary or bound state particles and their
antiparticles. In the case of CPT violation only in the mass term,
besides the difference in the masses of elementary bound state particles
and their antiparticles, there will be also an extremely tiny difference
in the lifetimes of bound states due to the difference in their phase
spaces. From the details of calculations, it appears that the separate
effects of the CPT violation described above are quite general, neither
depending on how the nonlocality is achieved, nor depending on what this
violation is due to: due to T violation, as considered in the present
work, which can be attributed to a cosmological direction of time; to CP
or to both T and CP violations. The latter two cases satisfy the
Sakharov's conditions for explaining the baryon asymmetry in the
Universe.
\end{abstract}

\vspace{.5em}
\noindent
The Lorentz invariance, CPT and spin-statistics theorems are key
properties of any relativistic quantum field theory (QFT) and
understanding their relations remains a fundamental issue.
We seek to answer the question what does violation of CPT invariance
lead to in a Lorentz invariant QFT?

According to the CPT theorem, any local Lorentz invariant QFT is CPT
invariant \cite{Luders:1954zz,Pauli:1955}.
Furthermore, the CPT theorem and the spin-statistics theorem are among
the few general results which can be proven in axiomatic quantum field
theory, without reference to a particular Lagrangian or Hamiltonian
model \cite{Jost:1957zz,Streater:1964,Jost:1965,Bogoliubov:1973,
Bogoliubov:1990,Alvarez-Gaume:2012zen,Nishijima:2023}. Crucial
consequences of the CPT theorem include the equality of the masses, the
decay widths (lifetimes) and the magnetic moments of a particle and its
antiparticle, which hold for both elementary and composite particles.
No violation of the CPT invariance
has been observed in experiments so far. Considering the fundamental
role of CPT invariance in QFT, it is no wonder that
there has been a common belief that relativistic quantum field theories
are necessarily CPT invariant, and further on that violation of CPT
invariance would necessarily imply violation of Lorentz invariance,
although neither of the said beliefs have been proven. There is no proof
that the Wightman axioms can be satisfied for every interacting theory
in four dimensions. Indeed the axiomatic framework does not yet cover
(non-Abelian) gauge theories or interacting theories in general
\cite{Streater:1964,Jost:1965,Bogoliubov:1973,Bogoliubov:1990}.
Hence the standing of CPT invariance in relativistic QFT is a pertinent
problem \cite{Chaichian:2011fc,Duetsch:2012sd}.

Since the possibility of CPT invariance violation has mainly been
studied in theoretical frameworks where Lorentz invariance is also
violated, it is important to study the consequences of CPT violation
in a Lorentz invariant theory. That is necessary for uncovering what CPT
violation alone entails.

In this letter, we consider relativistic QFT where CPT invariance is
broken but Lorentz invariance is valid.
It has been shown that CPT violation does not lead to violation of
Lorentz invariance and vice versa \cite{Chaichian:2011fc}. That was
achieved by introducing nonlocal interactions and/or nonlocal mass
terms, which are Lorentz invariant but violate CPT invariance
\cite{Chaichian:2011fc,Chaichian:2012ga,Chaichian:2012hy}. Such theories
can be considered as effective theories, which can be used to explore
the consequences of CPT violation that may arise in some fundamental
theory.

The other direction of the relation between Lorentz invariance and CPT
invariance, i.e. that violation of Lorentz invariance does not lead to
CPT violation, has been known longer, since QFT on
noncommutative spacetime preserves CPT invariance
\cite{Sheikh-Jabbari:2000hdt,Chaichian:2002vw,Alvarez-Gaume:2003lup,
Chaichian:2004qk}, while the Lorentz invariance
is broken or more precisely deformed as a part of the twisted
Poincar\'e symmetry
\cite{Chaichian:2004za,Wess:2003da,Chaichian:2004yh}. We should
also note that the spin-statistics relation remains intact in the
noncommutative QFT \cite{Chaichian:2002vw} based on the
Groenewold-Moyal star product, $f(x) \star g(x) = f(x) \exp\left(
\frac{i}{2}\overleftarrow{\partial}_\mu \theta^{\mu\nu}
\overrightarrow{\partial}_\nu \right) g(x)$.
A general proof has been given in \cite{Tureanu:2006pb} for the validity
of spin-statistics relation in noncommutative QFT, based on twisted
Poincar\'e symmetry discovered in \cite{Chaichian:2004za}.
Note that in a deformed QFT based on the representations of
$\kappa$-deformed Poincar\'e algebra, which is Lorentz violating,
however, the spin-statistics relation is lost, and the notion of
symmetrized or antisymmetrized multiparticle states ceases to have
meaning \cite{Arzano:2022vmh}.

The main conclusion of the Letter is that CPT violation has two separate
consequences, depending on whether Lorentz invariant CPT violation
is implemented in an interaction term or in a mass term:
\begin{enumerate}
\item If only a CPT-violating interaction term is included, the masses
of an elementary particle and its antiparticle remain equal but the
equalities of their lifetimes and cross sections are broken.
Furthermore, the equalities of the masses and the lifetimes of bound
states and their anti-states can be broken.
\item If CPT violation is implemented only in a mass term of the field,
what implies that the particle is not pointlike, the equality of the
masses is broken, while equalities of cross sections and magnetic
moments remain intact, but with a tiny difference in lifetimes due to
slight effect of phase spaces.
\end{enumerate}
If both CPT noninvariant interactions and mass terms are included,
their effects are combined: the equalities of the masses, the decay
widths and the magnetic moments of an elementary particle and its
antiparticle can all be broken. The situation with composite particles
is more involved. If a difference in the masses of a composite
particle and its antiparticle is observed, it could be due to either
CPT-violating mass terms or CPT-violating interactions or both, in its
constituent particles. All these consequences have to be taken into
account in any experimental test of CPT invariance, e.g. in tests
involving possible differences in the properties of matter and
antimatter. On top of that, we emphasize that the aforementioned
consequences do not imply violation of Lorentz invariance.

It is also important to understand the relation of the CPT invariance
and the spin-statistics theorem. The spin-statistics relation
\cite{Pauli:1940zz,Luders:1958zz,Burgoyne:1958}, and the Pauli exclusion
principle as its manifestation, are essential for the structure and
stability of matter \cite{Lieb:1975mi,Lieb:2005}.
In axiomatic QFT the requirements for the
spin-statistics theorem are stricter than the requirements for the CPT
invariance: the proof of the spin-statistics theorem requires that the
Wightman functions are invariant under cyclic permutations of fields at
Jost spacetime points, while the proof of the spin-statistic theorem
requires that the Wightman functions are invariant under any permutation
of fields at Jost spacetime points \cite{Jost:1957zz,Streater:1964,
Jost:1965,Bogoliubov:1973,Bogoliubov:1990}. Therefore,
we could expect that the violation of CPT invariance must imply
violation of the spin-statistics relation. However, we must keep in mind
that the axiomatic framework might not be fully applicable here, since
the present theory involves nonlocal terms, and also since even a local
interacting theory might not satisfy the axioms, as remarked above.
We show that the spin-statistics theorem remains valid in the theory of
Lorentz invariant CPT violation.

\paragraph{Lorentz invariant CPT violation}
Lorentz invariant CPT violation can be realized with a nonlocal
(interaction) term that includes the factor $\theta(x^0-y^0)$, where
$\theta$ is the Heaviside step function and $x^0$, $y^0$ are the time
coordinates of two points in spacetime. Another factor is included to
ensure Lorentz invariance, which also defines the scope of the nonlocal
interaction: e.g. $\theta((x-y)^2)$ for all causally connected points
\cite{Chaichian:2011fc} or $\delta((x-y)^2-l^2)$ for all points with a
certain spacetime interval \cite{Chaichian:2012ga}, where $l$ is a real
constant parameter. Another possible choice for the second factor is
$\theta((x-y)^2)e^{-(x-y)^2/l^2}$ \cite{Chaichian:2011fc}, where the
real parameter $l$ is the scale of the nonlocal interaction. The product
of those two factors, e.g. $\theta(x^0-y^0)\delta((x-y)^2-l^2)$, ensures
the invariance under proper orthochronous Lorentz transformations, since
the order of the times $x^0$ and $y^0$ is unchanged for timelike
intervals and the second factor ensures that spacelike intervals do not
contribute to the interaction.

Quantization of theories which are nonlocal in time is challenging,
since canonical quantization cannot be relied on due to the lack of a
reliable definition of canonical momenta. Therefore, we do not yet have
a perfectly satisfying technique for the quantization of theories that
are nonlocal in time. The path integral quantization based on
Schwinger's action principle has been applied to the nonlocal
CPT-violating interactions with interesting results
\cite{Chaichian:2012ga}. This approach follows the same lines as the
successful path integral quantization of space-time noncommutative
theories \cite{Fujikawa:2004rt}. Hence we quantize the theory with the
path integral formulation based on Schwinger's action principle.

\paragraph{CPT-violating interactions}
Various types of CPT noninvariant interactions can be considered by
introducing the aforementioned nonlocal factor into an interaction term.
Some examples include self-interactions like a nonlocal modification of
$\lambda\phi^4$ theory and interactions between fields of different
types. Since all the elementary interactions of matter are described
with interactions between fermions and bosons, we show here an example
of such an interaction. As a nonlocal CPT-violating interaction between
a spin-$1/2$ field $\psi$ and a real scalar field $\phi$ we consider a
Yukawa-type interaction. An interaction vertex is introduced into the
Lagrangian as in \cite{Chaichian:2012ga} (see also
\cite{Chaichian:2011fc}):
\begin{equation}\label{Yukawa-type}
\mathcal{L}_Y=g\bar\psi(x)\psi(x)\phi(x) +g_1\bar\psi(x)\psi(x)\int d^4y
\theta(x^0-y^0)\delta((x-y)^2-l^2)\phi(y),
\end{equation}
where the nonlocal term with the coupling constant $g_1$ is odd under
T and CPT but retains CP invariance. The effect of the
CPT-violating interaction in momentum space can be represented with the
following two \emph{form factors}
\begin{equation}\label{formfactors}
f_{\pm}(k)=\int d^4z\, e^{\pm ik\cdot z}\theta(z^0)\delta(z^2-l^2),
\end{equation}
which are related as $f_\pm(-k)=f_\mp(k)$. For a timelike momentum $k$
we may choose a Lorentz frame such that $\vec{k}=0$, and then the form
factors can be written as
\begin{equation}\label{formfactors2}
f_{\pm}(k^0)=2\pi\int_0^\infty dz\frac{z^2e^{\pm ik^0\sqrt{z^2+l^2}}}
{\sqrt{z^2+l^2}}
\overset{(k^0>0)}{=}\frac{2\pi}{(k^0)^2}\int_0^\infty dz\frac{z^2e^{\pm
i\sqrt{z^2+(k^0l)^2}}}
{\sqrt{z^2+(k^0l)^2}},
\end{equation}
where $k^0>0$ is assumed in the second equality.
Form factors are related to the size of particles (and in the
case of a composite particle also to the internal structure of the
particles).

Consider the consequences of the inclusion of the CPT-violating
interaction in the tree level processes (compared to the local
Yukawa interaction, i.e. just the first term in \eqref{Yukawa-type}).
The amplitude of the fermion pair creation process
$\phi\rightarrow\bar\psi\psi$ is changed so that the Yukawa coupling
constant $g$ is replaced by $g+g_1 f_+(k)$, where $k$ is the
four-momentum of the scalar particle. The amplitude of the fermion
annihilation process $\bar\psi\psi\rightarrow\phi$ is changed so that
$g$ is replaced by $g+g_1f_-(k)$. Consequently, the amplitudes for
these two processes, which are related via time reversal, contain
different phases due to the two form factors \eqref{formfactors}. There
is a similar result in the amplitude of the process
$\phi\phi\rightarrow\bar\psi\psi$, since $g^2$ is replaced by
$(g+g_1f_+(k_1))(g+g_1f_+(k_2))$, where $k_1$ and $k_2$ are the momenta
of the incoming scalar particles, while in the amplitude of the process
$\bar\psi\psi\rightarrow\phi\phi$, $g^2$ is replaced by
$(g+g_1f_-(k_1))(g+g_1f_-(k_2))$.
Such differences in the phases result to differences in decay and
scattering amplitudes (as well as in induced dipole moments).
The absolute squares of the tree level amplitudes do not feature the
T-violating phases, since $(f_\pm(k))^{*}=f_\mp(k)$. Hence for example
the aforementioned tree level amplitudes give $|g+g_1
f_+(k)|^2=|g+g_1 f_-(k)|^2$ and
$|(g+g_1f_+(k_1))(g+g_1f_+(k_2))|^2=|(g+g_1f_-(k_1))(g+g_1f_-(k_2))|^2$.
However, when loop corrections are included, the T-violating phases of
the tree level contributions interfere with the phases of the loop
corrections, which results into detectable differences in the probabilities
of the two processes that are related via time reversal. For example,
the one-loop $O(g^3)$ contributions to the amplitudes of the processes
$\phi\rightarrow\bar\psi\psi$ and $\bar\psi\psi\rightarrow\phi$ have the
same phase, say $e^{-i\alpha}$, while the tree-level $O(g_1)$
contributions have different phases, $e^{\pm i\alpha_{\mathrm{CPT}}}$.
The interference of such phases leads to the violation of time reversal
and CPT invariances in the squared decay and scattering amplitudes
(for detailed discussion, see \cite{Chaichian:2012ga}).

Note that some scattering amplitudes for the interaction
\eqref{Yukawa-type} do not exhibit CPT violation effect at tree
level. E.g., consider the amplitude of elastic scattering of fermion and
antifermion, $\bar\psi\psi\rightarrow\bar\psi\psi$. The $s$-channel
contribution of the Yukawa interaction is changed so that $g^2$ is
replaced by $(g+g_1f_+(p_1+p_2))(g+g_1f_-(p_1+p_2))$, while in the
$t$-channel contribution $g^2$ is replaced by
$(g+g_1f_+(p_1-p_3))(g+g_1f_-(p_1-p_3))$, which contain only symmetric
combinations of the form factors \eqref{formfactors}. For this process
one needs to look into loop corrections with higher powers of $g_1$.

The fermion self-energy corrections do not introduce splitting of the
fermion and antifermion masses. The theory has residual symmetries
under C and CP, which is sufficient to maintain the equality of the
fermion and antifermion masses. The one-loop fermion self-energy
correction for the interaction \eqref{Yukawa-type} is obtained as
\begin{multline}
\Sigma(p)=(2\pi)^4\delta^{(4)}(p'-p)\int\frac{d^4k}{(2\pi)^4}
\frac{1}{\slashed{p}-\slashed{k}-m}\frac{1}{k^2-M^2}\\ \times\left[
g^2+gg_1\left(f_+(k)+f_-(k)\right)+g_1^2f_+(k)f_-(k) \right],
\end{multline}
where $m$ and $M$ are the fermion and scalar particle masses,
respectively. The self-energy corrections contain symmetric (and
Hermitian) combinations of the form factors \eqref{formfactors} such as
$f_+(k)f_-(k)$ and $f_+(k)+f_-(k)$, which are symmetric under
$k\rightarrow-k$ and do not break the symmetry between positive and
negative fermion energy $p_0$. Thus, there is no splitting in fermion
and antifermion masses. This result is retained even in the presence of
CP violation \cite{Chaichian:2012ga}, when the local Yukawa interaction
is replaced by $g\bar\psi(x)(1+i\varepsilon\gamma_5)\psi(x)\phi(x)$
with a small real $\varepsilon$, which indicates that the invariance
under C is sufficient to maintain the equality of the masses.

While we have considered a CPT-violating Yukawa interaction in the
leading order, the main conclusion remains valid in general and
including all higher order corrections, as shown below by involving the
gauge invariance arguments.

In gauge theory, gauge invariance together with Lorentz invariance
ensures that the mass of a fermion ($m$) is equal to the mass of its
antiparticle ($\bar m$). For simplicity, consider quantum
electrodynamics. Due to the gauge invariance, replacing the photon
$A_\mu$ in the fermion-photon vertex $e\bar\psi\gamma^\mu A_\mu\psi$,
with $\partial_\mu$ (in $x$-space) or with $k_\mu$ in momentum space
does not change the vertex, i.e. it gives zero: $\bar\psi\gamma^\mu
k_\mu\psi=0\Rightarrow m=\bar m$; $k=p_\psi-p_{\bar\psi}$. The
electromagnetic current is conserved, as depicted by the Ward identity.

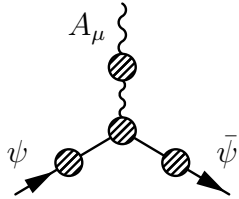
\begin{figure}[ht]
\begin{center}
\begin{fmffile}{QED-vertex-cor}
\begin{fmfgraph*}(80,80)
\fmftop{t1} \fmfbottom{b1,b2}
\fmf{photon,label=$A_\mu$}{t1,v1}
\fmf{photon}{v1,v2}
\fmf{fermion,label=$\psi$,label.side=left}{b1,v3}
\fmf{plain}{v3,v2}
\fmf{plain}{v2,v4}
\fmf{fermion,label=$\bar\psi$,label.side=left}{v4,b2}
\fmfblob{10}{v1,v2,v3,v4}
\end{fmfgraph*}
\end{fmffile}
\end{center}
\caption{The hashed circles stand for all the quantum corrections to the
fermion-photon interaction vertex.}
\label{fig}
\end{figure}

Now consider a theory with nonlocal interactions, where the action is
still invariant under the gauge transformation \cite{Chaichian:2012hy}:
$\psi(x)\rightarrow e^{i\alpha(x)}\psi(x)$,
$A_\mu(x)\rightarrow A_\mu(x)+\frac{1}{e}\partial_\mu\alpha(x)$,
with the gauge field coupled to all charged fields.
The free Lagrangian is taken to be the local one.
Consider merely for simplicity that only a Yukawa interaction is
taken to be nonlocal \eqref{Yukawa-type}.
Corrections to the fermion-photon vertex are depicted in
Fig.~\ref{fig}, where the dashed circles denote the corrections of all
orders to the propagators of $A_\mu$, $\psi$ and $\bar\psi$, and to the
vertex. The gauge invariance still ensures that the renormalized masses are equal,
$m=\bar m$.

\paragraph{Splitting of the masses of an elementary particle
and its antiparticle}
If one starts with the general Hermitian CPT-invariant Lagrangian of a
free fermion and modifies any of its terms to become nonlocal by
introducing a nonlocal factor with the aforementioned method, no
splitting between the masses and the widths of the particle and its
anti-particle appears \cite{Chaichian:2012ga}.
However, while in a local theory the term $i\mu\bar\psi(x)\psi(x)$ with 
real $\mu$ cancels with its Hermitian conjugate, a nonlocal term like
$i\mu F(x,y)\bar\psi(x)\psi(y)$, where $F(x,y)$ is the nonlocal factor,
e.g. $\theta(x^0-y^0)\delta((x-y)^2-l^2)$, gives a nonvanishing 
Hermitian contribution to the action. Thus, we consider the action
\cite{Chaichian:2012bk}
\begin{equation}\label{S.masssplitting}
\begin{split}
S&=\int d^{4}x\left\{\bar{\psi}(x)i\gamma^{\mu}\partial_{\mu}\psi(x)
 - m\bar{\psi}(x)\psi(x) \right.\\
&\quad -\int
d^{4}y\left.\left[\theta(x^{0}-y^{0})-\theta(y^{0}-x^{0})\right]
\delta((x-y)^2-l^2)[i\mu\bar{\psi}(x)\psi(y)]\right\},
\end{split}
\end{equation}
where for a real parameter $\mu$ the nonlocal term is odd under charge
conjugation and even under parity and time reversal.
Hence the action \eqref{S.masssplitting} is odd under C, CP and CPT.

Dirac equation in the momentum space, with the Ansatz
$\psi(x)=e^{ip\cdot x}\chi(p)$, is
\begin{equation}
\slashed{p}\chi(p)=m\chi(p)+i\mu\left[f_{+}(p)-f_{-}(p)
\right]\chi(p),
\end{equation}
where the two form factors $f_\pm(p)$ are defined in
\eqref{formfactors}.
The Lorentz covariant off-shell propagator is defined by
\begin{equation}\label{propagator.masssplitting}
\int d^{4}x\,d^{4}y\,
e^{ip\cdot(x-y)}\braket{T^{\star}\psi(x)\bar{\psi}(y)}
=\frac{i}{\slashed{p}-m+i\epsilon-i\mu[f_{+}(p)-f_{-}(p)]},
\end{equation}
where the poles are shifted due to the form factors. The poles occur
only for timelike momentum. The eigenvalue equation in the rest frame
($\vec{p}=0$) is
\begin{equation}\label{p0.eigenvalue}
p_0=\gamma_0\left( m-4\pi\mu\int_{0}^{\infty}dz\frac{z^2\sin
[p_0\sqrt{z^2+l^2}]}{\sqrt{z^2+l^2}} \right).
\end{equation}
The CPT transformed eigenvalue equation, which is obtained by
$p_0\rightarrow-p_0$ and sandwiching with $\gamma_5$, has a similar
form but the sign of the second term on the right-hand side is opposite,
\begin{equation}\label{p0.eigenvalue2}
p_0=\gamma_0\left( m+4\pi\mu\int_{0}^{\infty}dz\frac{z^2\sin
[p_0\sqrt{z^2+l^2}]}{\sqrt{z^2+l^2}} \right).
\end{equation}
For $\mu\gg m$, we can estimate the masses of the particle and its
antiparticle by solving the eigenvalue equations iteratively to the
first order in $\mu$: \eqref{p0.eigenvalue} and \eqref{p0.eigenvalue2}
give the mass eigenvalues as
\begin{equation}\label{m_mp}
m_\mp\approx m\mp4\pi\mu\int_{0}^{\infty}dz\frac{z^2\sin
[m\sqrt{z^2+l^2}]}{\sqrt{z^2+l^2}}.
\end{equation}
Thus, the difference of the form factors \eqref{formfactors2} in
\eqref{p0.eigenvalue}, i.e. $f_+(p_0)-f_-(p_0)$, results into the
splitting of the masses of the particle and its antiparticle.

The additional mass term in the action \eqref{S.masssplitting}
contributes to the free propagator \eqref{propagator.masssplitting} of
the field but does not contribute to any interaction vertices.
Therefore, the equality of the widths and cross sections for a particle
and its antiparticle is not altered by the CPT noninvariant mass term.

\paragraph{Spin-statistics theorem}
The classic form of the spin-statistics theorem was established by Pauli
\cite{Pauli:1940zz} and his proof involved the CPT invariance as one of
the assumptions. The spin-statistics theorem was proven without assuming
the CPT theorem by L\"uders and Zumino \cite{Luders:1958zz} and Burgoyne
\cite{Burgoyne:1958}. Those traditional proofs of the spin-statistics
relation rely on the operator formalism.
The spin-statistics relation can be proven also in the path integral
formalism \cite{Fujikawa:2001vy}.

We present key parts of the proof in the theory with Lorentz invariant
CPT violation. Consider the CPT-violating interaction
\eqref{Yukawa-type}. The free fields are unaltered and hence the proofs
of \cite{Fujikawa:2001vy} apply for them. The off-shell propagators for
the interacting fields are obtained from the path integral as
\begin{equation}\label{propagator.spin1/2}
\braket{T^{\star}\psi(x)\bar{\psi}(y)}
=\frac{i}{i\slashed{\partial}_x-m+g\phi(x)+g_1\int d^4z
F(x,z)\phi(z)+i\epsilon}\delta^4(x-y),
\end{equation}
where $F(x,y)=\theta(x^0-y^0)\delta((x-y)^2-l^2)$, and
\begin{equation}\label{propagator.spin0}
\braket{T^{\star}\phi(x)\phi(y)}
=\frac{i}{\Box_x+M^2+g\bar{\psi}(x)\psi(x)+g_1\int d^4z
F(z,x)\bar{\psi}(z)\psi(z)-i\epsilon}\delta^4(x-y).
\end{equation}
The Bjorken--Johnson--Low (BJL) method
\cite{Bjorken:1966jh,Johnson:1966se}
enables the definition of the $T$-product in terms of the
$T^{\star}$-product, defined for the propagator
\eqref{propagator.spin1/2} as
\begin{equation}\label{T-product}
\begin{split}
\int d^{4}x\,e^{ip\cdot(x-y)}\braket{T\psi(x)\bar{\psi}(y)}
&=\int d^{4}x\,e^{ip\cdot(x-y)}\braket{T^{\star}\psi(x)\bar{\psi}(y)}\\
&-\lim_{p_0\rightarrow0}\int
d^{4}x\,e^{ip\cdot(x-y)}\braket{T^{\star}\psi(x)\bar{\psi}(y)}.
\end{split}
\end{equation}
From such two-point correlation functions the equal time commutation
and anticommutation relations for the fields are derived
\cite{Fujikawa:2001vy}.
Assuming that the path integral measure for the field $\psi$
is valued in Grassmann variables, we obtain with the BJL method from
\eqref{propagator.spin1/2} and \eqref{T-product} that
\begin{equation}
\delta(x^0-y^0)\braket{\{\psi(x),\bar{\psi}(y)\}}=\gamma^0\delta^4(x-y),
\end{equation}
which is the basic anticommutation relation for the spin-$1/2$ field,
as well as that $\int d^{4}x\,e^{ip\cdot(x-y)}\braket{T\left(
i\slashed{\partial}_x-m+g\phi(x)+g_1\int d^4z
F(x,z)\phi(z) \right)\psi(x)\bar{\psi}(y)}=0$, which is consistent with
the field equation for $\psi$. Both $\psi$ and $\bar\psi$ anticommute
with themselves, i.e.
$\delta(x^0-y^0)\braket{\{\psi(x),\psi(y)\}}=0$ and
$\delta(x^0-y^0)\braket{\{\bar{\psi}(x),\bar{\psi}(y)\}}=0$, since
the corresponding propagators vanish,
$\braket{T^{\star}\psi(x)\psi(y)}=0$ and
$\braket{T^{\star}\bar{\psi}(x)\bar{\psi}(y)}=0$.
The positive energy condition is satisfied with the Feynman prescription
$m-i\epsilon$. The norm in the Hilbert space is positive definite only
when the spin-$1/2$ field in the path integral is Grassmann
variable.
The scalar field $\phi$ is considered with the same method using the
propagator \eqref{propagator.spin0}.

Similar derivation is performed for the case of the CPT-violating theory
with particle-antiparticle mass splitting \eqref{S.masssplitting} using
the propagator \eqref{propagator.masssplitting}. In this case the
derivation is similar to the usual free field case, since the form
factors in the propagator vanish in the equal-time limit,
$[f_+(p)-f_-(p)]\rightarrow0$ when $p_0\rightarrow\infty$.
Therefore, the spin-statistics theorem remains valid
when CPT-violating interactions and mass terms are present.
It is interesting that a similar result occurs in the case of
infinite-component QFT, where CPT can be violated, while spin-statistics
theorem remains intact \cite{Oksak:1968aj}. Seemingly,
infinite-component QFT theories in a way mimic the effect of nonlocality
as is the case in string theories.

As a side remark to reach still another picture of
quantization for CPT-violating but Lorentz invariant QFT, it could be
beneficial to mention a quantization method introduced by Umezawa and
Takahashi \cite{Takahashi:1964zz}, which was based on Umezawa’s unique
approach to the quantization of local relativistic field theories
\cite{Umezawa:1956,Takahashi:1969,Umezawa:1993}. This method does not
require canonical formulation and it might be possible to generalize it
for nonlocal theories like the present one.

As final remarks, we would like to mention a few important results,
which emphasize general aspects of CPT violation:
\begin{itemize}
\item Quantum corrections due to interaction terms always shift the
values of the masses. However, the changes in the mass are equal for the
particle and its antiparticle in all Lorentz invariant theories, where
the mass term is not nonlocal. Therefore, only a nonlocal and
CPT-violating mass term, such as the last term in
\eqref{S.masssplitting}, which corresponds to the form-factor, i.e. to a
non-point-like particle, can give different masses for the particle and
its antiparticle.

\item From the details and performing the calculations in this work, it
has appeared to us within a certain degree of assurance that the
results and the effects of different CPT-violating terms obtained here
are quite general ones and do not depend on how the nonlocality is
achieved, and as well do not depend on how the CPT violation has
appeared:
\begin{itemize}
\item Nonlocal QFT can be constructed in several ways, namely by
a) the Schwinger point like prescription; b) by smearing of the field
operators, as presented in the present work; c) by using the higher
derivative theories, such as the ones already used for many years and
highlighted in some works, such as in \cite{Biswas:2014yia} and
\cite{Abu-Ajamieh:2023syy}, and therein for references to previous works
on the subject in the literature.

\item CPT violation in the nonlocal QFT can be achieved in a few ways:
i) as in the present work by breaking the time reversal, what could be
attributed to the cosmological arrow of time, where time evolves only
towards future, on which many works have appeared. Examples can be
found, e.g. in
\cite{Hawking:1987na,Sakharov:1980kc,Barnaveli:1993np,Bowman:2014fca}
and references therein; ii) by breaking the CP invariance, introducing a
complex phase, originating e.g. due to the experimentally verified
Kobayashi--Maskawa quark mixing, while keeping the time invariance; also
iii) by breaking both the CP invariance and independently also breaking
the time invariance. The two cases ii) and iii), whereby both CP and C
violation are present is one of the conditions for the Sakharov
mechanism \cite{Sakharov:1967dj} to explain the matter-antimatter
asymmetry in Universe.
\end{itemize}
\end{itemize}
We see that there are still several further possibilities to be
considered together with their consequent implications on observable
effects, some of which are under study and will be presented in a future
communication.

\paragraph{Acknowledgements}
We are grateful to Michael D\"utsch, Jos\'e Gracia-Bondia, Adam
Schwimmer, Ivan Todorov and Anthony Zee for several illuminating
discussions. Our special thanks go to Kazuo Fujikawa for contributing to
this work through several previous works on CPT-violating nonlocal QFT.

\newcommand{\atitle}[1]{\emph{#1},}
\newcommand{\booktitle}[1]{\emph{#1}}
\newcommand{\ebooktitle}[2]{\href{https://doi.org/#2}{\emph{#1}}}
\newcommand{\jref}[2]{\href{https://doi.org/#2}{#1}}
\newcommand{\arXiv}[2]{\href{http://arxiv.org/abs/#1}
{\texttt{arXiv:#1 [#2]}}}
\newcommand{\arXivOld}[1]{\href{http://arxiv.org/abs/#1}
{\texttt{arXiv:#1}}}

\end{document}